\documentclass[floatfix,aps,prb,showpacs,twocolumn,floatfix,superscriptaddress]{revtex4}
\usepackage{bm}
\usepackage{epsfig}
\usepackage{amssymb}
\usepackage{graphicx}
\usepackage{amsmath}
\usepackage[sort&compress]{natbib}     
\usepackage{graphicx}         
\usepackage{color}
\usepackage{amsmath}
\usepackage{enumerate}
\newcommand{\shpar}{\shortparallel}

\begin{document}

\title{Quantitative modeling of spin relaxation in quantum dots} 

\author{J. P. Hansen}
\affiliation{Department of Physics and Technology, University
  of Bergen, N-5007 Bergen, Norway}
\affiliation{Nanoscience Center, Department of Physics, 
University of Jyv\"askyl\"a, P.O. Box 35, FI-40014 Jyv\"askyl\"a, Finland}

\author{S. A. S\o rng\aa rd}
\affiliation{Department of Physics and Technology, University
  of Bergen, N-5007 Bergen, Norway}

\author{M. F\o rre }
  \affiliation{Department of Physics and Technology, University
  of Bergen, N-5007 Bergen, Norway}

\author{E. R\"as\"anen}
\affiliation{Nanoscience Center, Department of Physics, 
University of Jyv\"askyl\"a, P.O. Box 35, FI-40014 Jyv\"askyl\"a, Finland}
\affiliation{Physics Department, Harvard University, 02138 Cambridge MA, USA}


\date{\today}

\begin{abstract}
We use numerically exact diagonalization to calculate the 
spin-orbit and phonon-induced triplet-singlet relaxation rate 
in a two-electron quantum dot exposed to a tilted magnetic field. 
Our scheme includes a three-dimensional description of the
quantum dot, the Rashba and the linear and cubic Dresselhaus 
spin-orbit coupling, the ellipticity of the quantum dot, 
and the full angular description of the magnetic field.
We are able to find reasonable agreement with the
experimental results of Meunier {\it et al.} 
[Phys. Rev. Lett. {\bf 98}, 126601 (2007)] in terms of
the singlet-triplet energy splitting and the spin relaxation rate, 
respectively. We analyze in detail the effects of the spin-orbit
factors, magnetic-field angles, and the dimensionality,
and discuss the origins of the remaining deviations
from the experimental data.
\end{abstract}

\maketitle

\section{Introduction}

Quantum dots (QDs) are well-known examples of confined and 
quantized systems in semiconductor heterostructures. Among other 
potential applications in, e.g., quantum optics, QDs have been
experimentally realized as controllable quantum bits.~\cite{petta}
On the road to quantum computing applications it is essential 
to understand the electronic and spin-dependent relaxation 
processes in QDs.~\cite{hanson,Fujisawa}

The coupling between quantized states and lattice vibrations 
in semiconductor structures has been examined already for 
decades.~\cite{Meijer1953, Zook1964,Leburton}
In QDs the spin relaxation has attracted attention
both 
theoretically~\cite{Chaney_2007,Khaetskii_2001,Golovach2004,Florescu_2006,
Climente_2007_Rapid,golovach:045328,Climente2007,Dickmann_2007} and
experimentally.~\cite{Nature3,Nature4,Nature2,Nature1,Hanson_2005,Meunier}
Recently, Meunier {\it et al.}~\cite{Meunier} 
measured the triplet-singlet relaxation rate in a 
two-electron (quasi) two-dimensional (2D) QD. 
They also derived a semi-empirical model reproducing the 
general trends of the experimental data. However,
a good agreement was found using a spin-orbit (SO) coupling
that depended on the magnetic field, and that was
significantly smaller than the SO strength reported in 
earlier studies.~\cite{zumbuhl}

In this work we describe a two-electron QD
using both 2D and three-dimensional (3D) models, respectively, 
and include (i) the description of a tilted external magnetic 
field, (ii) the ellipticity of the 
QD, and (iii) both the Rashba and the linear and cubic Dresselhaus 
spin-orbit (SO) interaction. We point out that
none of the previous theoretical approaches have aimed at
such a complete description, which is shown to be important
when comparing with experiment at the most detailed level.
The two-electron states are calculated 
numerically exactly by diagonalizing the Hamiltonian,
and the phonon-induced relaxation rate is obtained from
Fermi's golden rule. We find a reasonable agreement with
the experimental results,~\cite{Meunier} provided that the 3D
description and a moderate ellipticity of the QD are applied. 
The remaining discrepancy  between experiment and theory can
be caused by a more complex dot geometry of the 
experiment, and/or it indicates the limit of 
the widely used effective mass approximation 
for electrons confined by a harmonic potential.

\section{Model and method}\label{model}

Our two-electron QD exposed to an external magnetic field ${\bf B}$ 
is described by the Hamiltonian
\begin{equation}
H= \sum_{i=1,2}
\big[h({\mathbf{r}}_i)+h_{\rm SO}({\mathbf{r}}_i) \big]
+ \frac{e^2}{4 \pi \epsilon_\textrm{r} \epsilon_0 \left| \mathbf{r}_1-\mathbf{r}_2 
\right|},
\label{eq1}
\end{equation}
where $\mathbf{r}_i=(x_i,y_i,z_i)$ with $i=1,2$ are the 
coordinates of the two electrons. 
The single-electron Hamiltonian
is given by a sum of the kinetic, external potential, and Zeeman 
terms as
\begin{eqnarray}
h & = & -\frac{\hbar^2}{2m^*}\left(\nabla+e{\mathbf A}\right)^2 \\
& + & \frac{1}{2}m^*\left[\delta\omega_0 x^2+(1/\delta)\omega_0 y^2+\omega_z z^2\right]+g^*\mu_\textrm{B} {\mathbf B}\cdot{\mathbf S}. \nonumber 
\label{ham}
\end{eqnarray}
We consider magnetic fields tilted from the {\em xy} 
plane with an angle $\theta$, and tilted azimuthally
from the {\em y} axis with an angle $\phi$. Thus the magnetic
field has an expression
\begin{equation}
{\mathbf B}=B_0(\cos\phi\sin\theta,\sin\phi\sin\theta,\cos\theta). 
\end{equation}
The vector potential in  the symmetric gauge, 
${\mathbf A} = \left( {\mathbf B} \times {\mathbf r}\right)/2 $, 
then becomes 
\begin{eqnarray}
{\mathbf A} & = & \frac{B_0}{2}(z\sin\theta\sin\phi -y \cos\theta, \nonumber \\ 
& & x \cos\theta -z \cos\phi\sin\theta, \nonumber \\
& & y\cos\phi\sin\theta -x \sin\phi\sin\theta),
\end{eqnarray}
which approaches 
${\mathbf A}={B_0}(-y, x)/2$ 
in the 2D limit ($z\rightarrow 0$).

The harmonic confinement potential is asymmetric both laterally and
vertically: the ellipticity in the {\em xy} plane can be tuned
with the $\delta$-parameter, and the in-plane and off-plane 
confinement strengths are fixed to $\hbar\omega_0=2.5$ meV 
and $\hbar\omega_z=11.9$ meV, respectively. The material
parameters are chosen to be those of GaAs, i.e., effective mass $m^*=0.067m_\textrm{e}$,
relative permittivity $\epsilon_\textrm{r}=12.4$ and the gyromagnetic ratio 
$g^*=-0.44$. The constant $\mu_\textrm{B}=e/(2 m_\textrm{e})$ in Eq.~(\ref{ham}) 
is the Bohr magneton.

A harmonic form has been shown to describe well the 
electron confinement in both lateral and vertical 
semiconductor (GaAs-type) quantum dots.~\cite{reimann} 
The model was established soon after the first
Coulomb-blockade experiments, when the measured
and calculated addition energies were compared.~\cite{kouwenhoven}
More recently, the harmonic model has been shown
to explicitly yield the measured single-eletron 
spectrum,~\cite{impurity} and also the spin-blockade
oscillations for both lateral and vertical QD devices 
up to about 50 electrons.~\cite{spindroplet,rogge}

The SO interaction in Eq.~(\ref{eq1}) is given by
\begin{equation}
	h_{\rm SO}= h_{R} + h_D 	
\end{equation}
with the Rashba term
\begin{equation}
h_{R} = \frac{\alpha}{\hbar} (p_x \sigma^y - p_y\sigma^x )
\end{equation}
and the Dresselhaus term~\cite{stano} 
containing both linear and cubic contributions,
\begin{eqnarray}
	h_D &=& \frac{\gamma}{\hbar^3} \langle p_z^2 \rangle (p_y\sigma^y - p_x\sigma^x) \nonumber \\ 
& + & \frac{\gamma}{2\hbar^3} (p_x p_y^2 \sigma^x - p_y p_x^2 \sigma^y).
\end{eqnarray}
The coupling strengths of the Rashba and Dresselhaus terms
are set by parameters $\alpha$ and $\gamma$, respectively,  
and $\sigma^x$ and $\sigma^y$ are the Pauli matrices.
We choose $\alpha=0$ and $\gamma=11.15\,{\rm eV{\AA}^3}$
as our default values,
but consider also different magnitudes 
to assess the sensitivity 
of the spin relaxation to the SO coupling. 
The measurements of Meunier {\em et al.}~\cite{Meunier} and 2D 
calculations~\cite{Climente2007,krich} suggest relatively
small coupling constants. 
This is in agreement with the experiment of Zumb\"u{}hl {\it et al.}~\cite{Zumb05} reporting $\gamma=9\, {\rm eV \AA^3}$ which is notably smaller than the commonly used value of $27.5\, {\rm eV\AA^3}$.

The total wave function is a product of the $xy$ (planar) 
component $\psi_\shpar({\mathbf{r}}_1,{\mathbf{r}}_2)$, 
obtained from exact diagonalization of the Hamiltonian in Eq.~(\ref{eq1})
in a (large) basis of spin-symmetrized products of
Cartesian single-electron harmonic-oscillator 
eigenstates,~\cite{harjuPRL,Popsueva} and a
simple  $z$  (perpendicular) component $\psi_z(z_1,z_2)$
incorporating the material thickness.
The $z$ component of the wave function is conveniently 
modeled in terms of the ground state
wave function of a  harmonic potential, with confinement 
strength $ \omega_z $ defining an effective
confinement length, 
$\mathcal{L}_z = \sqrt{4\hbar/(m^* \omega_z)}$, 
in the $z$ direction.

The spectrum is obtained by diagonalizing the Hamiltonian 
in a basis of spin symmetrized
two-electron harmonic-oscillator states. Basis sizes  
$|n_x, n_y, n_z\rangle $ up to $n_{max} = 6$ 
for the $(x,y)$ coordinates and $n_z=0$
for the vertical coordinate are included. 
Analytical matrix elements are then obtained for
all terms except for the electron-electron interaction 
in Eq. (\ref{eq1}). 

It was shown by Popsueva
{\em et al.}~\cite{Popsueva} that,  in the 2D case with
$\omega_x=\omega_y=\omega_0$,  
all matrix elements can be evaluated analytically.
In particular, for any double pairs of basis functions ($i,j$) 
the four-dimensional integrals over the electron-electron
interaction can be expressed as
\begin{equation}
I_{\rm 2D}^{r_{12}} = \omega_0^{1/2} \int_{-\infty}^{\infty} F_{ij}^{\rm 2D}(s_{\shpar}) ds_{{\shpar}},
\label{r122D}
\end{equation}
where 
\begin{equation}
F_{ij}^{\rm 2D}(s_{\shpar}) = \sum_n a_n s_{{\shpar}}^{2n} e^{-s_{{\shpar}}^2}.
\end{equation}
The variable $s_{\shpar}$ is here defined from the Bethe integral,
\begin{equation}
\frac{1}{|\mathbf{r}_1 - \mathbf{r}_2|}  = \frac{1}{2\pi^{2}} \int  \frac{d^3 s}{s^2} e^{i \mathbf{s} \cdot \mathbf{r}_1} e^{-i \mathbf{s} \cdot \mathbf{r}_2}.
\end{equation}
When this expression is introduced in calculating the matrix 
element between any pair of basis functions
$(i,j)$, the expression above is derived 
with $s_{\shpar} = \sqrt{s_x^2 + s_y^2}$.
When the integrals have been analytically evaluated once, 
the matrix elements for any confinement $\omega_0$ are 
obtained by a simple multiplication. 

In 3D the electron-electron repulsion amounts to 
more complicated six-dimensional integrals. 
When extending with harmonic confinement
also in the $z$ direction a similar scaling 
relation can be obtained, i.e.,
\begin{eqnarray}
I_{\rm 3D}^{r_{12}} &=& \omega_0^{1/2} \int_{-\infty}^{\infty} F_{ij}^{\rm 2D}(s_{\shpar}) 
\left[ 1 - {\rm erf}\left(\sqrt{\frac{1}{2\omega_z}}s_{{\shpar}}\right) \right] \nonumber \\
& \times & e^{\frac{s_{{\shpar}}^2}{2\omega_z}} ds_{{\shpar}}
\label{r123D}
\end{eqnarray}
Here ${\rm erf}(x) =  2/\sqrt{\pi}\int_0^x \exp(-t^2) dt $ 
is the error function. 
The matrix elements for the electron-electron 
interaction can thus be obtained for any $\omega_0$ by
scaling as in 2D when they have been calculated 
for a given value of the vertical confinement.
We note also that the formula above provides 
a continuous route from 3D to 2D, since
\begin{equation}
I_{\rm 2D}^{r_{12}} = \lim_{\omega_z \rightarrow \infty} I_{\rm 3D}^{r_{12}}(\omega_z).
\end{equation}
For 2D calculations the finite thickness can be 
consistently taken into account by replacing
Eq. (\ref{r122D}) by Eq. (\ref{r123D}) 
when calculating the interaction terms.

The triplet-singlet relaxation rate is calculated using 
Fermi's golden rule, which -- in the
case of phonon emission -- may be written as,~\cite{Bockelmann}
\begin{eqnarray}
	\Gamma &=& \frac{V}{\hbar (2\pi)^2}  \sum_{j=1}^3 \int d^3 q |M_j(\mathbf{q})|^2 
 \nonumber \\
 &\times&  \left| \langle S | \widehat{H}_\textrm{ph}  | T \rangle \right|^2 
 \delta\left( \Delta E -\hbar c_jq\right), 
\label{eq:rate}
\end{eqnarray}
where $T$ and $S$ refer to the initial singlet and final 
triplet states, respectively,
$V$ is the normalization volume, 
$\Delta E$ is the singlet-triplet energy splitting, 
$\mathbf{q}=(\textbf{q}_\shpar,q_z)$ is the momentum of the released phonon, and
$\widehat{H}_\textrm{ph} = \sum_{i=1,2} e^{-i(\textbf{q}_\shpar\cdot \textbf{r}_i  +q_z z_i )}$
is the phonon coupling.
Linear dispersion relations
have been adopted, i.e., we set 
$ \omega_\textrm{S,T}\equiv\Delta E/\hbar=c_\textrm{l,t} q$.
Here
$c_\textrm{l}=4720$ m/s ($c_\textrm{t}=3340$ m/s) denotes the 
longitudinal (transverse) speed of sound.~\cite{Meunier} 
The electrons couple to longitudinal acoustic 
phonons through the deformation potential
coupling~\cite{Climente06,Ridley1982}
\begin{equation}
	\left |M_1(\mathbf{q})\right|^2= \frac{\hbar\Xi_\textrm{d}^2}{2\rho c_\textrm{l} V}|\mathbf{q}|,
	\label{eq32}
\end{equation}
as well as the piezoelectric coupling,
\begin{equation}
	\left |M_2(\mathbf{q})\right|^2= 
	\frac{32\pi^2\hbar e^2 h_{14}^2}{\epsilon_r^2\rho c_\textrm{l} V} \frac{(3q_xq_yq_z)^2}{|\mathbf{q}|^7},
	\label{}
\end{equation}
whereas coupling to transverse
phonons only takes place through the latter one,
\begin{equation}
	\left |M_3(\mathbf{q})\right|^2 \!\!=\!\! 
	\frac{32\pi^2\hbar e^2h_{14}^2}{\epsilon_\textrm{r}^2\rho c_\textrm{t} V}\!\!
	\left(\!\frac{q_x^2q_y^2+q_y^2q_z^2+q_z^2q_x^2}
	{|\mathbf{q}|^5}   
	\!-\! \frac{(3q_xq_yq_z)^2}{|\mathbf{q}|^7}\! \right).
	\label{eq31}
\end{equation}
In these expressions $\rho=5300$ kg/m$^3$ is the GaAs mass density,
$\Xi_\textrm{d}=6.7$ eV is the deformation potential constant, and
$h_{14}=1.4\times 10^9$ V/m is the piezoelectric constant.
There are two transverse phonon modes, and hence 
$M_3(\mathbf{q})$ is considered twice when computing the rate.

The matrix element $ \langle S | \widehat{H}_\textrm{ph}  | T \rangle$ 
in Eq.~(\ref{eq:rate})
separates into a product of a planar ($xy$) and a perpendicular 
($z$) component, i.e.,
\begin{eqnarray}
\nonumber
 \langle S | \widehat{H}_\textrm{ph}  | T \rangle &\!=\!&
 \sum_{i=1,2}  \langle \psi_\shpar^{(S)}  |  e^{-i {\textbf{q}_\shpar \cdot \textbf{r}}_i} | 
 \psi_\shpar^{(T)} \rangle\langle {\psi_z} | e^{-i q_zz_i} | {\psi_z} \rangle\\
\nonumber
 &\!=\!&
2\!\!\sum_{k,l,k'}\!\! a_{k,l}^* a_{k',l}  \Phi_{n_k,n_{k'}} 
\Phi_{m_k,m_{k'}}\exp\left(-\frac{\hbar q_z^2}{4m^{*}\omega_z}\right). \\
\label{eq21}                             
\end{eqnarray}
Here $| \psi_\shpar^{(S)}\rangle$ and $| \psi_\shpar^{(T)}\rangle$ 
are the planar two-electron singlet and
triplet states, respectively. We have assumed 
that the $z$ component of the wave function is frozen,
i.e., no excitations are considered in that direction.
The expansion coefficients $a_{k,l}$ are obtained from the
diagonalization of the Hamiltonian in Eq.~(\ref{eq1}).
$n_k$ and $m_k$ are the 
single-particle harmonic quantum numbers
of one of the electrons  in the  $x$ and $y$ direction, 
respectively. The factor of two in Eq.~(\ref{eq21}) arises from the 
 fact that the total wave function of each state is antisymmetric. 
 The single-particle matrix elements 
 $\Phi_{n_i,n_{i'}}=\langle\phi_{n_i}|e^{-iq_xx}|\phi_{n_{i'}}\rangle$ 
are essentially associated Laguerre polynomials in 
$q^2$ multiplied by a damping exponential function, 
\begin{equation}
\label{eq23}
\Phi_{n,n'} \!=\!  (- i)^\eta \frac{N!}{\sqrt{n!n'!}} \tilde{q}_x^{2M+\eta}
	\exp\!\big(\!-\tilde{q}_x^2/2\big)
	L_{N}^{2M+\eta}(\tilde{q}_x^2).
\end{equation}
Here $\tilde{q}_x=\hbar q_x/(2m^* \omega_0)^{1/2}$, $N = \min(n,n')$, $M=(|n-n'|-\eta)/2$, 
and $\eta=1$ if $n+n'$ is odd and 0 otherwise (analogously in $y$)~\cite{sorngard}. 
The phonon induced  triplet-singlet relaxation rate is now calculated by 
inserting Eqs.~(\ref{eq32})-(\ref{eq21}) into Eq.~(\ref{eq:rate}) and performing
the integration over the phonon momentum $\textbf{q}$.

\section{Results}

\subsection{Singlet-triplet energy splitting}

\begin{figure}
\includegraphics[width=0.8\columnwidth]{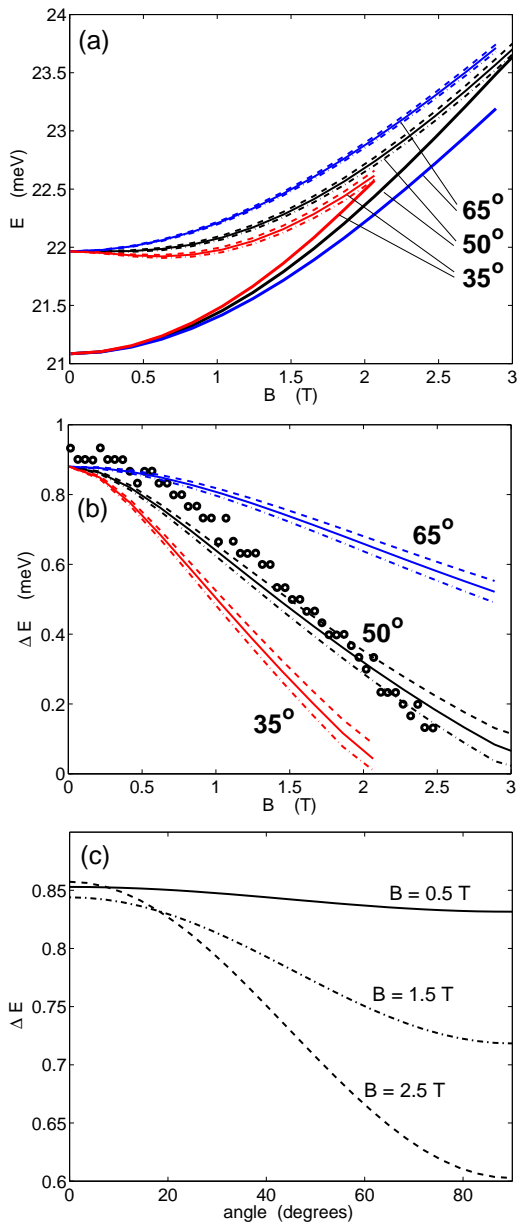}
\caption{ 
(Color online)  
(a) Energy of singlet ground state (thick lines) and 
three lowest triplet states (thin lines) 
in a two-electron harmonic
($\hbar\omega_0= 2.5$ meV, $\hbar\omega_z = 11.9$ meV), elliptic 
($\delta=1.3$) quantum dot as function of the magnetic
field strength for three different tilting angles of the magnetic field, 
$\theta=65^\circ$, $50^\circ$, and $35^\circ$. In the triplet states the dashed,
solid, and dash-dotted lines indicate $S_z=-1,\,0,$ and $+1$, respectively.
(b) Singlet-triplet energy splitting for the three cases of the upper panel. 
Open circles correspond to the experimental data in Ref.~\onlinecite{Meunier}. 
(c) Variation of $\Delta E (B)$ as a 
function of $\phi$ for $B=0.5$ T (solid line), 1.5 T (dash-dotted line), 
and 2.5 T (dashed line), respectively. The results for $B=1.5$ T and 2.5 T
have been multiplied by factors 1.4 and 2.4, respectively, to set the scale.
}
\label{fig1}
\end{figure}

Figure~\ref{fig1}(a) shows the total energies of singlet 
(thick lines) and triplet 
states (thin lines) in a 
two-electron, elliptic QD as a function of the magnetic field. 
The ellipticity of the QD is fixed to $\delta=1.3$ [see Eq.~(\ref{ham})].
The triplet states with $S_z=0,\pm 1$ are split into three curves due 
to the Zeeman effect which is linear in $B$. We show results for three
tilting angles of the magnetic field, $\theta=65^\circ$ (outermost lines), 
$50^\circ$ (middle lines), and $35^\circ$ (innermost lines) as
indicated in the figure.

The tilting angle $\theta$ has a relatively large effect on the energies, so that
the closing point of the singlet-triplet energy gap $\Delta E$ moves to 
larger magnetic fields when the angle is increased. Figure~\ref{fig1}(b) shows
the gap as a function of $B$ in comparison with the experimental 
values (circles). First, the downward ``bending'' of $\Delta E$ at small
fields is due to the ellipticity; in a circular QD the gap increases 
linearly as $B$ approaches zero. In this respect, we find a good agreement
with the experiment, and may conclude that the ellipticity of the real
QD device is close to our chosen value $\delta=1.3$.
Secondly, the best overall agreement is found with a tilting angle 
$\theta=50^\circ$ which is smaller than $\theta_\textrm{exp}=68^\circ \pm 5^\circ$ reported 
in Ref.~\onlinecite{Meunier}. 

Another angle affecting the energy gap is 
the azimuthal direction of the magnetic
field with respect to the axis of the ellipticity.  
In Figure~\ref{fig1}(c) we plot
the variation of $\Delta E (B)$ 
as a function of the azimuthal angle $\phi$ for 
three magnetic field strengths. We observe that 
the variation increases with increasing magnetic field, so that
close to the crossing ($B \sim 3T$) the variation is $\sim 20\%$. 
Combining this observation with Fig.~\ref{fig1}(b) leads to
a conclusion that the best agreement with the experimental
result is obtained with $\phi=90^\circ$, i.e., when the shortest
axis of the ellipse is parallel to the tilting direction
of the magnetic field. In the following we set $\phi=90^\circ$
unless stated otherwise.

With a fixed $\delta$ obtained from the bending as 
$B\rightarrow 0$ [see Fig.~\ref{fig1}(b)], the singlet-triplet
energy gap essentially depends on $\omega_0$ and $\omega_z$. 
SO parameters are so small that they play only a minor role in the 
energy gap. The parameter values used here, $\hbar\omega_0=2.5$ meV  and 
$\hbar\omega_z=11.9$ meV, give a reasonable agreement with the
experiment in the energy separation and in the spin-relaxation 
rate (see below). An increase in {\em both} parameters and
thus smaller vertical extension can also lead to 
good agreement in the energy gap, but a poor agreement in
the spin relaxation rate. This effect is further illuminated
in Sec. III.C.

We point out that the results are not sensitive to
using other types of frozen ground-state vertical basis 
functions and/or a larger number of vertical basis functions. 
The latter, as well as representing the vertical dimension by an 
infinite barrier at $z=0$ and a harmonic confinement
for $z>0$ does not lead to significant changes. We conclude that 
good agreement with the experiment is obtained for a narrow range 
of harmonic confinement parameters.
However, the theoretical magnetic tilting angle needs to be reduced to 
$\theta=50^\circ$ to
arrive to this level of agreement. The origin of this discrepancy pose
a challenge for future experiments and theory. In the remaining part of
this work we will stick to $\theta= 50^\circ$.

\subsection{Spin relaxation rate}

\begin{figure}
\begin{center}
\includegraphics[width=0.8\columnwidth]{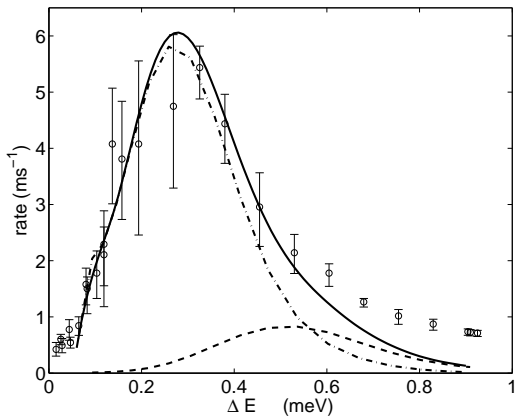}
\caption{
Calculated total triplet-singlet relaxation rate when the
magnetic field tilting angle is
$\theta=50^\circ$ (solid line) in comparison with the
experimental data in Ref.~\cite{Meunier} 
(circles with error bars).
The other parameters are the same as 
in Fig.~\ref{fig1}.
The solid line shows the total rate. 
The dashed and dash-dotted lines
show the contribution from the piezoelectric and
deformation potential coupling,
respectively. The spin-orbit coupling
parameters are $\gamma=11.15\,{\rm eV \AA^3}$ and $\alpha=0$
for the Dresselhaus and Rashba terms, respectively.
}
\label{fig2}
\end{center}
\end{figure}

In Fig.~\ref{fig2} we show the calculated total relaxation rate 
(solid line) in comparison with the experimental result (circles with 
error bars) in Ref.~\onlinecite{Meunier}.
Dashed and dash-dotted lines correspond to the contributions
from the piezoelectric coupling and the deformation potential 
coupling, respectively. We have used 
$\gamma=11.15\, {\rm eV  \AA^3}$ and $\alpha=0$ for the 
Dresselhaus and Rashba SO coupling strengths, respectively. 
Apart from the smallest and largest energy gaps, where 
the relaxation rate is very small, we obtain
a very good agreement with the experimental measurements.
We find that interchanging the values 
between the two couplings has only a minor effect on the
transition rate; in fact, this is 
true for any combination with 
$\sqrt{\alpha^2 + (\gamma \langle p_z^2 \rangle)^2} = 0.58\, {\rm meV \AA}$.  
The experiment of Meunier {\em et al.}~\cite{Meunier}
gives a measure of this quantity but cannot resolve the 
relative magnitude of the
two terms. Measurements of angular resolved 
phonons~\cite{angular_phonons}
or external fields controlling the Rashba term are 
needed to address the comparison between theory
and experiment at a more detailed level.
 
In general, the coupling to
the phonon bath is strongest when the QD size matches with
the phonon wavelength.~\cite{hanson,Bockelmann,Meunier,golovach:045328}
The piezoelectric coupling (dashed line) is found to dominate the 
relaxation rate in Fig.~\ref{fig2} at small energy splittings 
($\Delta E\lesssim 0.5$ meV), whereas the deformation potential
coupling dominates at larger $\Delta E$. This is expected 
due to the $\sqrt{\Delta E}$ and $\sqrt{1/\Delta E}$ dependence
of these couplings, respectively. In the former case, the 
coupling occurs through slower transverse phonons, which yields
a peak at smaller $\Delta E$.

\begin{figure}
\includegraphics[width=0.9\columnwidth]{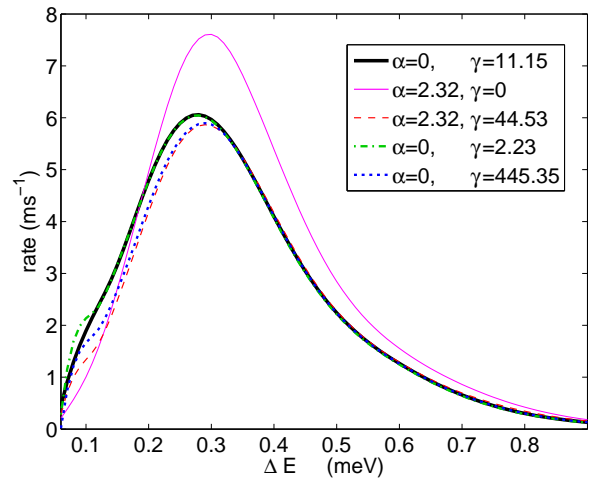}
\caption{(Color online) Scaled relaxation rates 
according to Eq.~(\ref{rate2}) 
for four different sets of spin-orbit coupling 
strengths. $\alpha$ and $\gamma$ are in units of ${\rm meV \AA}$ and $\rm eV \AA^3$, respectively.
} 
\label{fig3}
\end{figure}

Next we examine the effect of the SO coupling strength on
the relaxation rate in more detail. 
Figure~\ref{fig3} shows
the scaled relaxation rates for four different sets
of $\alpha$ and $\gamma$ in the units of ${\rm meV \AA}$ and ${\rm eV \AA^3}$, respectively. 
The black solid line corresponds to Fig.~\ref{fig2}.  
The other curves are divided by the actual $\alpha^2 + (\gamma\langle p_z^2 \rangle)^2 $ and
multiplied by the square of the reference value, $\gamma=11.15\,{\rm eV \AA^3}$.
Overall, no significant qualitative changes 
in the rate are obtained within a realistic parameter range.
In particular, the peak
position is insensitive to both $\gamma$ and 
$\alpha$, and thus we cannot achieve 
a better agreement with the experimental rate 
in Fig.~\ref{fig2} (circles) by tuning either 
the Dresselhaus or Rashba SO coupling, or both. 
Furthermore, the fact that all curves are very close 
to each other indicates that the transition 
rate scales to a good agreement as
\begin{equation}
\Gamma(\alpha, \gamma) \approx \frac{\left[\alpha^2 + (\gamma\langle p_z^2 \rangle)^2\right]}{\alpha_0^2 + (\gamma_0 \langle p_z^2\rangle)^2} \Gamma\left(\alpha_0, \gamma_0\right).
\label{rate2}
\end{equation}
The scaling formula can be derived from an interaction Hamiltonian 
$V_I \sim \alpha V_I^1 + \gamma V_I^2$ 
in the first-order perturbation limit when the 
interfering paths can be neglected, i.e.,
\begin{equation}
\langle f | V_I^1 | i \rangle \langle f | V_I^2 | i \rangle \approx 0.
\end{equation}
From Fig.~\ref{fig3} we see that the scaling for 
fixed $\theta,\phi$ is valid. This suggests that the spin relaxation
through the Rashba or Dresselhaus coupling in fact follows 
non-interfering paths within our basis states.

Generally, however, the SO coupling is anisotropic and depends on
the orientation of the magnetic field with respect to
the crystal axis.~\cite{hanson} Consequently, the relaxation rate is
anisotropic as well.~\cite{falko} Hence, in view of
a tilted magnetic field it is difficult to precisely 
assort the Rashba and Dresselhaus contributions 
in our QD device.

\begin{figure}
\includegraphics[width=0.99\columnwidth]{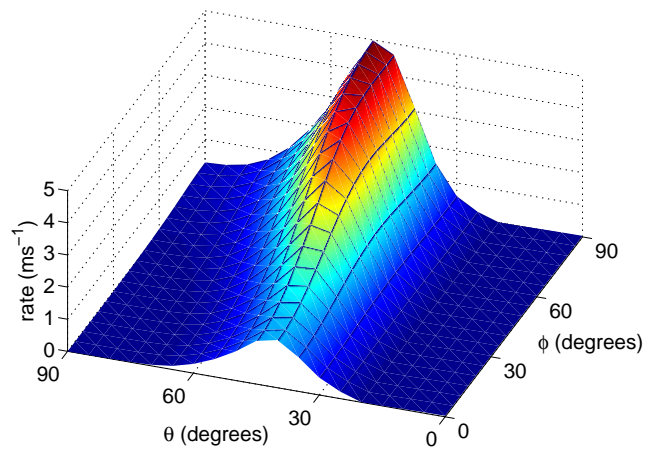}
\caption{(Color online)
Dependence of the relaxation rate on the 
magnetic-field angles $\theta$ and $\phi$.
The field strength is fixed to $B=2$ T.
The other parameters are the same as in Fig.~\ref{fig1}.
}
\label{fig4}
\end{figure}

In Fig.~\ref{fig4} we show how the relaxation rate for fixed 
magnetic field strength (here $B=2$ T) depends on the field 
direction with any combination of $\theta$ and $\phi$.
As pointed out above, the relaxation rate is strongly dependent
on the energy splitting and qualitatively follows a
form $\Delta E^n \exp{(-c \Delta E^2)}$ with positive constants 
$n$ and $c$. At $B=2$ T the rate is small for both $\theta \sim 0$
and $\theta \sim 90^\circ$, corresponding to small-$\Delta E$ and
large-$\Delta E$ limits in the qualitative formula, respectively.
At intermediate values of $\theta$ we find an area of
higher relaxation rates that increases as a function of $\phi$.
In this region of increased rates the energy splitting 
gives phonon wavelengths comparable with QD size.
Detailed analysis of the angular dependence
is not the scope of this work, but we finally 
point out that these results could be used as a guideline 
in the analysis of forthcoming experiments where the 
direction of the magnetic field can be varied;
in principle the angular properties could be 
used indirectly to obtain precise structural 
information of the QD device.

\subsection{Effects of dimensionality}

Finally we turn our attention to the effects 
of the dimensionality on the singlet-triplet energy
splitting and on the spin relaxation rate. 
We compare our 3D scheme as described in 
the beginning of Sec.~\ref{model} to a 
bare 2D approach, where only the in-plane confinement
and 2D basis functions are applied 
and the 2D approximation for the electron-electron
interaction is used (corresponding to $\omega_z \gg \omega_0$.  
This leads to relatively stronger electronic repulsion 
in the ground state than in the excited states, 
which in turn yields a smaller energy splitting.

\begin{figure}
\includegraphics[width=0.8\columnwidth]{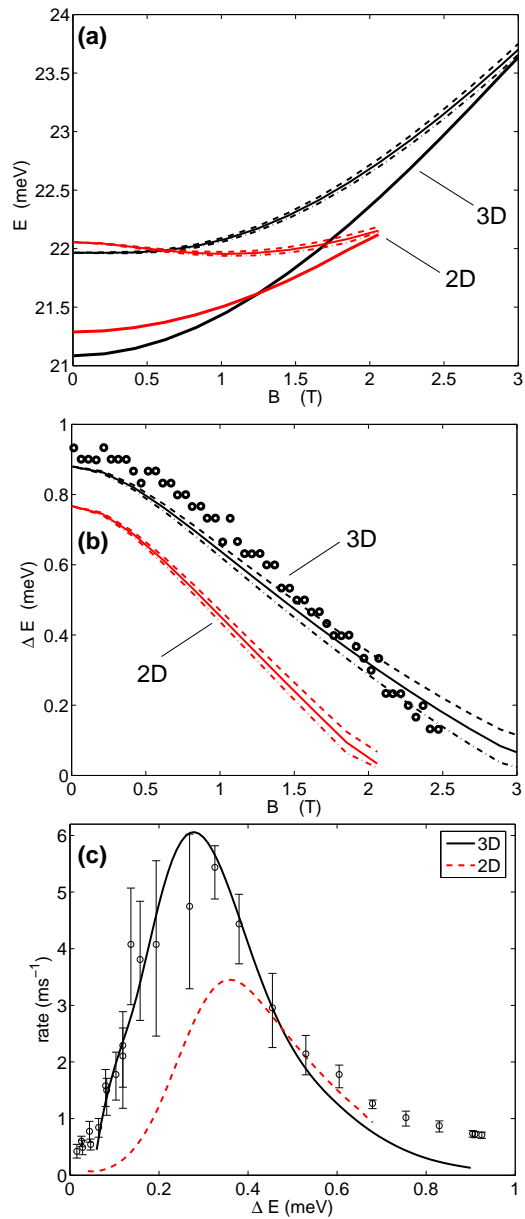}
\caption{(Color online) Comparison between two-dimensional 
and three-dimensional calculations, in both cases
with a tilted magnetic field of $\theta = 50^\circ$.  
(a) Energy levels of the lowest singlet
and triplet states. (b) Singlet-triplet energy
splitting. (c) Relaxation rate.
Parameters are the same as in Fig.~\ref{fig2}.
} 
\label{fig5}
\end{figure}

The results of the comparison are summarized in
Fig.~\ref{fig5}. First we notice that the tilted  
magnetic field has a very different effect on
the energy states in 3D and 2D (a).
In 2D the applied magnetic field
closes the energy gap already at $B\sim 2$ T, 
whereas in 3D the closing occurs at $B\sim 3$ T.
Therefore the 2D model cannot yield a reasonable
agreement with the experimental splitting as 
shown in Fig.~\ref{fig5}(b).
It should be noted that {\em if} the confinement strength
$\omega_0$ in the 2D model is slightly increased to yield
the correct splitting at $B=0$, the magnetic-field
dependence is still wrong, and the rate becomes much 
worse than the 3D result in comparison with the experiment.
Furthermore, as shown in Fig.~\ref{fig5}(c) 
the 2D model is unable to reproduce the experimental
spin relaxation rate. In contrast, the 3D model has 
a very good agreement with the experiment as
discussed above within Fig.~\ref{fig2}.

\section{Summary and discussion}

To summarize, we have proceeded towards quantitative
modeling of spin relaxation in semiconductor quantum dots
containing two electrons.
Our approach covers (i) a three-dimensional
description of the quantum dot device, (ii) numerically
exact treatment of the electron-electron interaction,
(iii) a tilted magnetic field, (iv) the ellipticity of the quantum dot, 
and (v) both the Rashba and the Dresselhaus spin-orbit coupling,
and the latter with both linear and cubic terms.
We have attributed the observed nonlinear behavior of the 
singlet-triplet energy gap at $B\lesssim 0.5$ T to
the ellipticity of the quantum dot. 
Then we have found a good agreement 
between theory and experiment in the
triplet-singlet energy splitting and in the relaxation 
rate, although the optimal tilting angle of the 
magnetic field was found to be considerably 
smaller than the experimental value.
Finally we have explicitly shown that
the three-dimensional model is essential to obtain
a reasonable agreement with the experiment 
in both the singlet-triplet energy splitting and 
the spin relaxation rate.
 
In view of our results supplied with 
aspects (i)-(v) listed above, the 
origins behind the remaining quantitative 
deviations between experiment and theory --
especially regarding the different angle of
the magnetic field -- is unclear. 
However, as the most 
probable scenario we may suggest that the actual 
confining potential deviates from harmonic, e.g., 
by being more strongly confined at the edge of the quantum 
dot. This can be induced by the gate geometry of
the lateral device and/or the nonuniformity of
the confinement in the vertical ($z$) direction.
These conditions can increase the high sensitivity
of the spectrum to the tilted magnetic 
field (Fig.~\ref{fig1}) even further.
In this respect, we hope that the present study 
motivates more efforts in this direction,
both theoretically and experimentally.

\begin{acknowledgments}
We thank L. M. K. Vandersypen and T. Meunier for valuable comments.
This work was supported by the Bergen Research
Foundation (Norway), the Research Council of Norway (RCN), the
Academy of Finland, the Wihuri Foundation, and the Nordforsk Network.
\end{acknowledgments}


\begin{thebibliography}{99}

\expandafter\ifx\csname natexlab\endcsname\relax\def\natexlab#1{#1}\fi
\expandafter\ifx\csname bibnamefont\endcsname\relax
  \def\bibnamefont#1{#1}\fi
\expandafter\ifx\csname bibfnamefont\endcsname\relax
  \def\bibfnamefont#1{#1}\fi
\expandafter\ifx\csname citenamefont\endcsname\relax
  \def\citenamefont#1{#1}\fi
\expandafter\ifx\csname url\endcsname\relax
  \def\url#1{\texttt{#1}}\fi
\expandafter\ifx\csname urlprefix\endcsname\relax\def\urlprefix{URL }\fi
\providecommand{\bibinfo}[2]{#2}
\providecommand{\eprint}[2][]{\url{#2}}

\bibitem[{\citenamefont{Petta et~al.}(2005)\citenamefont{Petta, Johnson,
  Taylor, Laird, Yacoby, Lukin, Marcus, Hanson, and Gossard}}]{petta}
\bibinfo{author}{\bibfnamefont{J.~R.} \bibnamefont{Petta}},
  \bibinfo{author}{\bibfnamefont{A.~C.} \bibnamefont{Johnson}},
  \bibinfo{author}{\bibfnamefont{J.~M.} \bibnamefont{Taylor}},
  \bibinfo{author}{\bibfnamefont{E.~A.} \bibnamefont{Laird}},
  \bibinfo{author}{\bibfnamefont{A.}~\bibnamefont{Yacoby}},
  \bibinfo{author}{\bibfnamefont{M.~D.} \bibnamefont{Lukin}},
  \bibinfo{author}{\bibfnamefont{C.~M.} \bibnamefont{Marcus}},
  \bibinfo{author}{\bibfnamefont{M.~P.} \bibnamefont{Hanson}},
  \bibnamefont{and} \bibinfo{author}{\bibfnamefont{A.~C.}
  \bibnamefont{Gossard}}, \bibinfo{journal}{Science}
  \textbf{\bibinfo{volume}{309}}, \bibinfo{pages}{2180} (\bibinfo{year}{2005}).

\bibitem[{\citenamefont{Hanson et~al.}(2007)\citenamefont{Hanson, Kouwenhoven,
  Petta, Tarucha, and Vandersypen}}]{hanson}
\bibinfo{author}{\bibfnamefont{R.}~\bibnamefont{Hanson}},
  \bibinfo{author}{\bibfnamefont{L.~P.} \bibnamefont{Kouwenhoven}},
  \bibinfo{author}{\bibfnamefont{J.~R.} \bibnamefont{Petta}},
  \bibinfo{author}{\bibfnamefont{S.}~\bibnamefont{Tarucha}}, \bibnamefont{and}
  \bibinfo{author}{\bibfnamefont{L.~M.~K.} \bibnamefont{Vandersypen}},
  \bibinfo{journal}{Rev. Mod. Phys.} \textbf{\bibinfo{volume}{79}},
  \bibinfo{eid}{1217} (\bibinfo{year}{2007}).

\bibitem[{\citenamefont{Fujisawa et~al.}(1998)\citenamefont{Fujisawa,
  Oosterkamp, van~der Wiel, Broer, Aguado, Tarucha, and
  Kouwenhoven}}]{Fujisawa}
\bibinfo{author}{\bibfnamefont{T.}~\bibnamefont{Fujisawa}},
  \bibinfo{author}{\bibfnamefont{T.~H.} \bibnamefont{Oosterkamp}},
  \bibinfo{author}{\bibfnamefont{W.~G.} \bibnamefont{van~der Wiel}},
  \bibinfo{author}{\bibfnamefont{B.~W.} \bibnamefont{Broer}},
  \bibinfo{author}{\bibfnamefont{R.}~\bibnamefont{Aguado}},
  \bibinfo{author}{\bibfnamefont{S.}~\bibnamefont{Tarucha}}, \bibnamefont{and}
  \bibinfo{author}{\bibfnamefont{L.~P.} \bibnamefont{Kouwenhoven}},
  \bibinfo{journal}{Science} \textbf{\bibinfo{volume}{282}},
  \bibinfo{pages}{932} (\bibinfo{year}{1998}).

\bibitem[{\citenamefont{Meijer and Polder}(1953)}]{Meijer1953}
\bibinfo{author}{\bibfnamefont{H.}~\bibnamefont{Meijer}} \bibnamefont{and}
  \bibinfo{author}{\bibfnamefont{D.}~\bibnamefont{Polder}},
  \bibinfo{journal}{Physica} \textbf{\bibinfo{volume}{19}}, \bibinfo{pages}{255
  } (\bibinfo{year}{1953}).

\bibitem[{\citenamefont{Zook}(1964)}]{Zook1964}
\bibinfo{author}{\bibfnamefont{J.~D.} \bibnamefont{Zook}},
  \bibinfo{journal}{Phys. Rev.} \textbf{\bibinfo{volume}{136}},
  \bibinfo{pages}{A869} (\bibinfo{year}{1964}).

\bibitem[{\citenamefont{Leburton et~al.}(1993)\citenamefont{Leburton, Pascual,
  and Sotomayor-Torres}}]{Leburton}
\bibinfo{author}{\bibfnamefont{J.~P.} \bibnamefont{Leburton}},
  \bibinfo{author}{\bibfnamefont{J.}~\bibnamefont{Pascual}}, \bibnamefont{and}
  \bibinfo{author}{\bibfnamefont{C.~M.} \bibnamefont{Sotomayor-Torres}},
  \emph{\bibinfo{title}{Phonons in Semiconductor Nanostructures}}
  (\bibinfo{publisher}{Kluwer Academic Publishers},
  \bibinfo{address}{Dordrecht}, \bibinfo{year}{1993}).


\bibitem[{\citenamefont{Golovach et~al.}(2008)\citenamefont{Golovach,
  Khaetskii, and Loss}}]{golovach:045328}
\bibinfo{author}{\bibfnamefont{V.~N.} \bibnamefont{Golovach}},
  \bibinfo{author}{\bibfnamefont{A.}~\bibnamefont{Khaetskii}},
  \bibnamefont{and} \bibinfo{author}{\bibfnamefont{D.}~\bibnamefont{Loss}},
  \bibinfo{journal}{Phys. Rev. B} \textbf{\bibinfo{volume}{77}},
  \bibinfo{pages}{045328} (\bibinfo{year}{2008}).

\bibitem[{\citenamefont{Climente
  et~al.}(2007{\natexlab{a}})\citenamefont{Climente, Bertoni, Goldoni, Rontani,
  and Molinari}}]{Climente2007}
\bibinfo{author}{\bibfnamefont{J.~I.} \bibnamefont{Climente}},
  \bibinfo{author}{\bibfnamefont{A.}~\bibnamefont{Bertoni}},
  \bibinfo{author}{\bibfnamefont{G.}~\bibnamefont{Goldoni}},
  \bibinfo{author}{\bibfnamefont{M.}~\bibnamefont{Rontani}}, \bibnamefont{and}
  \bibinfo{author}{\bibfnamefont{E.}~\bibnamefont{Molinari}},
  \bibinfo{journal}{Phys. Rev. B} \textbf{\bibinfo{volume}{76}},
  \bibinfo{pages}{085305} (\bibinfo{year}{2007}{\natexlab{a}}).

\bibitem[{\citenamefont{Golovach et~al.}(2004)\citenamefont{Golovach,
  Khaetskii, and Loss}}]{Golovach2004}
\bibinfo{author}{\bibfnamefont{V.~N.} \bibnamefont{Golovach}},
  \bibinfo{author}{\bibfnamefont{A.}~\bibnamefont{Khaetskii}},
  \bibnamefont{and} \bibinfo{author}{\bibfnamefont{D.}~\bibnamefont{Loss}},
  \bibinfo{journal}{Phys. Rev. Lett.} \textbf{\bibinfo{volume}{93}},
  \bibinfo{pages}{016601} (\bibinfo{year}{2004}).

\bibitem[{\citenamefont{Climente
  et~al.}(2007{\natexlab{b}})\citenamefont{Climente, Bertoni, Goldoni, Rontani,
  and Molinari}}]{Climente_2007_Rapid}
\bibinfo{author}{\bibfnamefont{J.~I.} \bibnamefont{Climente}},
  \bibinfo{author}{\bibfnamefont{A.}~\bibnamefont{Bertoni}},
  \bibinfo{author}{\bibfnamefont{G.}~\bibnamefont{Goldoni}},
  \bibinfo{author}{\bibfnamefont{M.}~\bibnamefont{Rontani}}, \bibnamefont{and}
  \bibinfo{author}{\bibfnamefont{E.}~\bibnamefont{Molinari}},
  \bibinfo{journal}{Phys. Rev. B} \textbf{\bibinfo{volume}{75}},
  \bibinfo{pages}{081303(R)} (\bibinfo{year}{2007}{\natexlab{b}}).

\bibitem[{\citenamefont{Khaetskii and Nazarov}(2001)}]{Khaetskii_2001}
\bibinfo{author}{\bibfnamefont{A.~V.} \bibnamefont{Khaetskii}}
  \bibnamefont{and} \bibinfo{author}{\bibfnamefont{Y.~V.}
  \bibnamefont{Nazarov}}, \bibinfo{journal}{Phys. Rev. B}
  \textbf{\bibinfo{volume}{64}}, \bibinfo{pages}{125316}
  (\bibinfo{year}{2001}).

\bibitem[{\citenamefont{Florescu and Hawrylak}(2006)}]{Florescu_2006}
\bibinfo{author}{\bibfnamefont{M.}~\bibnamefont{Florescu}} \bibnamefont{and}
  \bibinfo{author}{\bibfnamefont{P.}~\bibnamefont{Hawrylak}},
  \bibinfo{journal}{Phys. Rev. B} \textbf{\bibinfo{volume}{73}},
  \bibinfo{pages}{045304} (\bibinfo{year}{2006}).

\bibitem[{\citenamefont{Chaney and Maksym}(2007)}]{Chaney_2007}
\bibinfo{author}{\bibfnamefont{D.}~\bibnamefont{Chaney}} \bibnamefont{and}
  \bibinfo{author}{\bibfnamefont{P.~A.} \bibnamefont{Maksym}},
  \bibinfo{journal}{Phys. Rev. B} \textbf{\bibinfo{volume}{75}},
  \bibinfo{pages}{035323} (\bibinfo{year}{2007}).

\bibitem[{\citenamefont{Dickmann and Hawrylak}(2003)}]{Dickmann_2007}
\bibinfo{author}{\bibfnamefont{S.}~\bibnamefont{Dickmann}} \bibnamefont{and}
  \bibinfo{author}{\bibfnamefont{P.}~\bibnamefont{Hawrylak}},
  \bibinfo{journal}{J. Supercond.} \textbf{\bibinfo{volume}{16}},
  \bibinfo{pages}{387} (\bibinfo{year}{2003}).

\bibitem[{\citenamefont{Meunier et~al.}(2007)\citenamefont{Meunier, Vink, van
  Beveren, Tielrooij, Hanson, Koppens, Tranitz, Wegscheider, Kouwenhoven, and
  Vandersypen}}]{Meunier}
\bibinfo{author}{\bibfnamefont{T.}~\bibnamefont{Meunier}},
  \bibinfo{author}{\bibfnamefont{I.~T.} \bibnamefont{Vink}},
  \bibinfo{author}{\bibfnamefont{L.~H.} \bibnamefont{Willems van Beveren}},
  \bibinfo{author}{\bibfnamefont{K.-J.} \bibnamefont{Tielrooij}},
  \bibinfo{author}{\bibfnamefont{R.}~\bibnamefont{Hanson}},
  \bibinfo{author}{\bibfnamefont{F.~H.~L.} \bibnamefont{Koppens}},
  \bibinfo{author}{\bibfnamefont{H.~P.} \bibnamefont{Tranitz}},
  \bibinfo{author}{\bibfnamefont{W.}~\bibnamefont{Wegscheider}},
  \bibinfo{author}{\bibfnamefont{L.~P.} \bibnamefont{Kouwenhoven}},
  \bibnamefont{and} \bibinfo{author}{\bibfnamefont{L.~M.~K.}
  \bibnamefont{Vandersypen}}, \bibinfo{journal}{Phys. Rev. Lett.}
  \textbf{\bibinfo{volume}{98}}, \bibinfo{pages}{126601}
  (\bibinfo{year}{2007}).

\bibitem[{\citenamefont{Hanson et~al.}(2005)\citenamefont{Hanson, van Beveren,
  Vink, Elzerman, Naber, Koppens, Kouwenhoven, and Vandersypen}}]{Hanson_2005}
\bibinfo{author}{\bibfnamefont{R.}~\bibnamefont{Hanson}},
  \bibinfo{author}{\bibfnamefont{L.~H.} \bibnamefont{Willems van Beveren}},
  \bibinfo{author}{\bibfnamefont{I.~T.} \bibnamefont{Vink}},
  \bibinfo{author}{\bibfnamefont{J.~M.} \bibnamefont{Elzerman}},
  \bibinfo{author}{\bibfnamefont{W.~J.~M.} \bibnamefont{Naber}},
  \bibinfo{author}{\bibfnamefont{F.~H.~L.} \bibnamefont{Koppens}},
  \bibinfo{author}{\bibfnamefont{L.~P.} \bibnamefont{Kouwenhoven}},
  \bibnamefont{and} \bibinfo{author}{\bibfnamefont{L.~M.~K.}
  \bibnamefont{Vandersypen}}, \bibinfo{journal}{Phys. Rev. Lett.}
  \textbf{\bibinfo{volume}{94}}, \bibinfo{pages}{196802}
  (\bibinfo{year}{2005}).

\bibitem[{\citenamefont{Fujisawa et~al.}(2002)\citenamefont{Fujisawa, Austing,
  Tokura, Hirayama, and Tarucha}}]{Nature3}
\bibinfo{author}{\bibfnamefont{T.}~\bibnamefont{Fujisawa}},
  \bibinfo{author}{\bibfnamefont{D.~G.} \bibnamefont{Austing}},
  \bibinfo{author}{\bibfnamefont{Y.}~\bibnamefont{Tokura}},
  \bibinfo{author}{\bibfnamefont{Y.}~\bibnamefont{Hirayama}}, \bibnamefont{and}
  \bibinfo{author}{\bibfnamefont{S.}~\bibnamefont{Tarucha}},
  \bibinfo{journal}{Nature (London)} \textbf{\bibinfo{volume}{419}},
  \bibinfo{pages}{278} (\bibinfo{year}{2002}).

\bibitem[{\citenamefont{Elzerman et~al.}(2004)\citenamefont{Elzerman, Hanson,
  Willems~van Beveren, Witkamp, Vandersypen, and Kouwenhoven}}]{Nature4}
\bibinfo{author}{\bibfnamefont{J.~M.} \bibnamefont{Elzerman}},
  \bibinfo{author}{\bibfnamefont{R.}~\bibnamefont{Hanson}},
  \bibinfo{author}{\bibfnamefont{L.~H.} \bibnamefont{Willems~van Beveren}},
  \bibinfo{author}{\bibfnamefont{B.}~\bibnamefont{Witkamp}},
  \bibinfo{author}{\bibfnamefont{L.~M.~K.} \bibnamefont{Vandersypen}},
  \bibnamefont{and} \bibinfo{author}{\bibfnamefont{L.~P.}
  \bibnamefont{Kouwenhoven}}, \bibinfo{journal}{Nature (London)}
  \textbf{\bibinfo{volume}{430}}, \bibinfo{pages}{431} (\bibinfo{year}{2004}).

\bibitem[{\citenamefont{Kroutvar et~al.}(2004)\citenamefont{Kroutvar, Ducommun,
  Heiss, Bichler, Schuh, Abstreiter, and Finley}}]{Nature2}
\bibinfo{author}{\bibfnamefont{M.}~\bibnamefont{Kroutvar}},
  \bibinfo{author}{\bibfnamefont{Y.}~\bibnamefont{Ducommun}},
  \bibinfo{author}{\bibfnamefont{D.}~\bibnamefont{Heiss}},
  \bibinfo{author}{\bibfnamefont{M.}~\bibnamefont{Bichler}},
  \bibinfo{author}{\bibfnamefont{D.}~\bibnamefont{Schuh}},
  \bibinfo{author}{\bibfnamefont{G.}~\bibnamefont{Abstreiter}},
  \bibnamefont{and} \bibinfo{author}{\bibfnamefont{J.~J.}
  \bibnamefont{Finley}}, \bibinfo{journal}{Nature (London)}
  \textbf{\bibinfo{volume}{432}}, \bibinfo{pages}{81} (\bibinfo{year}{2004}).

\bibitem[{\citenamefont{Johnson et~al.}(2005)\citenamefont{Johnson, Petta,
  Taylor, Yacoby, Lukin, Marcus, Hanson, and Gossard}}]{Nature1}
\bibinfo{author}{\bibfnamefont{A.~C.} \bibnamefont{Johnson}},
  \bibinfo{author}{\bibfnamefont{J.~R.} \bibnamefont{Petta}},
  \bibinfo{author}{\bibfnamefont{J.~M.} \bibnamefont{Taylor}},
  \bibinfo{author}{\bibfnamefont{A.}~\bibnamefont{Yacoby}},
  \bibinfo{author}{\bibfnamefont{M.~D.} \bibnamefont{Lukin}},
  \bibinfo{author}{\bibfnamefont{C.~M.} \bibnamefont{Marcus}},
  \bibinfo{author}{\bibfnamefont{M.~P.} \bibnamefont{Hanson}},
  \bibnamefont{and} \bibinfo{author}{\bibfnamefont{A.~C.}
  \bibnamefont{Gossard}}, \bibinfo{journal}{Nature (London)}
  \textbf{\bibinfo{volume}{435}}, \bibinfo{pages}{925} (\bibinfo{year}{2005}).


\bibitem{zumbuhl} D. M. Zumb\"uhl, J. B. Miller, C. M. Marcus, K. Campman, and A. C. Gossard,
Phys. Rev. Lett. {\bf 89}, 276803 (2002).


\bibitem[{\citenamefont{Popsueva et~al.}(2007)\citenamefont{Popsueva, Nepstad,
  Birkeland, F\o{}rre, Hansen, Lindroth, and Waltersson}}]{Popsueva}
\bibinfo{author}{\bibfnamefont{V.}~\bibnamefont{Popsueva}},
  \bibinfo{author}{\bibfnamefont{R.}~\bibnamefont{Nepstad}},
  \bibinfo{author}{\bibfnamefont{T.}~\bibnamefont{Birkeland}},
  \bibinfo{author}{\bibfnamefont{M.}~\bibnamefont{F\o{}rre}},
  \bibinfo{author}{\bibfnamefont{J.~P.} \bibnamefont{Hansen}},
  \bibinfo{author}{\bibfnamefont{E.}~\bibnamefont{Lindroth}}, \bibnamefont{and}
  \bibinfo{author}{\bibfnamefont{E.}~\bibnamefont{Waltersson}},
  \bibinfo{journal}{Phys. Rev. B} \textbf{\bibinfo{volume}{76}},
  \bibinfo{pages}{035303} (\bibinfo{year}{2007}).

\bibitem{krich}  J. J. Krich and B. I. Halperin,
Phys. Rev. Lett. {\bf 98}, 226802 (2007).

\bibitem{Zumb05}
	D. M. Zumb\"uhl, J. B.  Miller,  C. M.  Marcus,  D. Goldhaber-Gordon, J. S. Harris, K. Campman and  A. C. Gossard
	Phys. Rev. B {\bf 72}, 081305 (2005) 

\bibitem{stano} P. Stano and J. Fabian, Phys. Rev. B {\bf 72}, 
155410 (2005).

\bibitem[{\citenamefont{Harju et~al.}(2002)\citenamefont{Harju, Siljam\"aki,
  and Nieminen}}]{harjuPRL}
\bibinfo{author}{\bibfnamefont{A.}~\bibnamefont{Harju}},
  \bibinfo{author}{\bibfnamefont{S.}~\bibnamefont{Siljam\"aki}},
  \bibnamefont{and} \bibinfo{author}{\bibfnamefont{R.~M.}
  \bibnamefont{Nieminen}}, \bibinfo{journal}{Phys. Rev. Lett.}
  \textbf{\bibinfo{volume}{88}}, \bibinfo{pages}{226804}
  (\bibinfo{year}{2002}).

\bibitem[{\citenamefont{Bockelmann}(1994)}]{Bockelmann}
\bibinfo{author}{\bibfnamefont{U.}~\bibnamefont{Bockelmann}},
  \bibinfo{journal}{Phys. Rev. B} \textbf{\bibinfo{volume}{50}},
  \bibinfo{pages}{17271} (\bibinfo{year}{1994}).

\bibitem[{\citenamefont{Climente et~al.}(2006)\citenamefont{Climente, Bertoni,
  Goldoni, and Molinari}}]{Climente06}
\bibinfo{author}{\bibfnamefont{J.~I.} \bibnamefont{Climente}},
  \bibinfo{author}{\bibfnamefont{A.}~\bibnamefont{Bertoni}},
  \bibinfo{author}{\bibfnamefont{G.}~\bibnamefont{Goldoni}}, \bibnamefont{and}
  \bibinfo{author}{\bibfnamefont{E.}~\bibnamefont{Molinari}},
  \bibinfo{journal}{Phys. Rev. B} \textbf{\bibinfo{volume}{74}},
  \bibinfo{eid}{035313} (\bibinfo{year}{2006}).

\bibitem[{\citenamefont{Ridley}(1982)}]{Ridley1982}
\bibinfo{author}{\bibfnamefont{B.~K.} \bibnamefont{Ridley}},
  \emph{\bibinfo{title}{Quantum processes in semiconductors}}
  (\bibinfo{publisher}{Clarendon Press}, \bibinfo{address}{Oxford},
  \bibinfo{year}{1982}).




\bibitem{angular_phonons} J. I. Climente, A. Bertoni, G. Goldoni, and E. 
Molinari, Phys. Rev. B {\bf 75}, 245330 (2007).

\bibitem{falko} V. I. Fal'ko, B. L. Altshuler, and O. Tsyplyatyev, 
Phys. Rev. Lett. {\bf 95}, 076603 (2005).

\bibitem{sorngard} S. A. S\o{}rngard, M. F\o{}rre and J. P. Hansen, New J. Phys., in print (2012).

\bibitem{reimann} S. M. Reimann and M. Manninen, Rev. Mod.
  Phys. {\bf 74}, 1283 (2002).

\bibitem{kouwenhoven} L. P. Kouwenhoven, D. G. Austing, 
and S. Tarucha, Rep. Prog. Phys. {\bf 64}, 701 (2001).

\bibitem{impurity} E. R\"as\"anen, J. K\"onemann, 
R. J. Haug, M. J. Puska, and R. M. Nieminen,
Phys. Rev. B {\bf 70}, 115308 (2004).

\bibitem{spindroplet} E. R\"as\"anen, H. Saarikoski, 
A. Harju, M. Ciorga, and A. S. Sachrajda,
Phys. Rev. B {\bf 77}, 041302(R) (2008).

\bibitem{rogge} M. C. Rogge, E. R\"as\"anen, and R. J. Haug,
Phys. Rev. Lett. {\bf 105}, 046802 (2010).

\end{thebibliography}
\end{document}